\documentclass[aps, 10pt, prl,twocolumn,amssymb,superscriptaddress,floatfix,endnotes]{revtex4-2}
\usepackage{graphicx}
\usepackage{amsmath}
\usepackage{amssymb}
\usepackage{color}
\usepackage[dvipsnames]{xcolor}
\usepackage{gensymb}
\usepackage{siunitx}
\usepackage{bm}
\usepackage[hidelinks]{hyperref}
\usepackage{verbatim}
\usepackage[normalem]{ulem}

\newcommand{\costh}{\ensuremath{\langle \cos^2 \theta_{\text{2D}} \rangle}}
\newcommand{\CSS}[0]{$\mathrm{CS_2}$}
\newcommand{\db}{\ensuremath{\Delta\beta}}
\newcommand{\dg}{\ensuremath{\Delta\gamma}}
\newcommand{\dt}{\ensuremath{\Delta t}}
\newcommand{\fcf}{\ensuremath{f_{\mathrm{cf}}}}
\newcommand{\fus}{\ensuremath{f_{\mathrm{us}}}}
\newcommand{\dfus}{\ensuremath{\Delta f_{\mathrm{us}}}}

\begin{document}

\title{An ultraslow optical centrifuge with arbitrarily low rotational acceleration}

\author{Kevin Wang}
\author{Ian MacPhail-Bartley}
\author{Cameron E. Peters}
\author{Valery Milner}
\email{vmilner@phas.ubc.ca}

\affiliation{Department of Physics \& Astronomy, The University of British Columbia, Vancouver, Canada}

\date{\today}

\begin{abstract}
We present the design and characterization of a laser pulse shaper, which creates an ``ultraslow optical centrifuge'' - a linearly polarized field whose polarization vector rotates with arbitrarily low angular acceleration. By directly recording this rotation in time with nonlinear cross-correlation, we demonstrate the tunability of the initial and final rotational frequencies of such a centrifuge in the range of accelerations, which are three orders of magnitude lower than those available with a conventional centrifuge design. We showcase the functionality of the ultraslow centrifuge by spinning \CSS{} molecules in a molecular jet. Utilizing the extremely low angular acceleration to control molecular rotation inside viscous media is a promising application for this unique optical tool.
\end{abstract}

\maketitle

\section{Introduction}
An optical centrifuge is an ultrafast laser-based instrument with numerous applications in molecular sciences (for recent reviews, see Refs.~\citenum{MacPhail2020, Mullin2025}). In a recent publication~\cite{Wang2025}, we argued that the ability of such a device to control molecular rotation is limited to molecules with relatively low moment of inertia and centrifugal distortion constant. The former sets an upper bound to the angular acceleration of the centrifuge, whereas the latter constrains the range of accessible rotational frequencies. By the very design of the original optical centrifuge~\cite{Karczmarek1999,Villeneuve2000}, it provided an efficient way to spin up molecules in the gas phase at a fast rate of $\sim \SI{100}{GHz/ps}$ to extremely high rotational states, known as ``super-rotors'', reaching frequencies of order $\SI{10}{THz}$.

On the other hand, the original design of the centrifuge makes it too fast to spin molecules embedded in strongly interacting environments, such as helium nanodroplets, where molecular rotors hold great potential as ``nanoprobes'' of superfluidity~\cite{Toennies2004, Stienkemeier2006, Choi2006, Slenczka2022}. Due to the weak bonding of helium atoms to the solvated molecule, both the effective moment of inertia and the centrifugal distortion constant of the ``molecule-He'' complex increase dramatically~\cite{Grebenev2000, Lehnig2009, Pentlehner2013, Chatterley2020, Cherepanov2021}. Hence, for a given torque exerted by the centrifuge field (whose maximum strength is capped by photo-ionization), the angular acceleration of the molecule appeared insufficient for keeping up with the rotating centrifuge, preventing rotational control inside He nanodroplets~\cite{Joergensen2018, MacPhail2024}.

An alternative approach to creating an optical centrifuge with zero angular acceleration by using two time-delayed frequency-chirped laser pulses of opposite circular polarization has been proposed theoretically~\cite{Faucher2018}, and used as a method of frequency shearing in a transparent dielectric medium for ultrashort pulse characterization~\cite{Bejot2020}. Applying this method to molecular ensembles, in Ref.~\cite{Wang2025} we demonstrated experimentally that such a ``constant-frequency'' centrifuge (cfCFG) is indeed capable of controlling molecular rotation in the desired low range of rotational frequencies around $\SI{10}{GHz}$, finally enabling us to optically centrifuge molecules embedded in helium nanodroplets~\cite{MacPhail2026}. At the same time, zero acceleration of cfCFG also meant that the enforced adiabatic rotational ladder climbing (typically expected from a conventional optical centrifuge) could not be executed, which resulted in a rather limited degree of rotational control.

With the goal of bringing back the adiabatic centrifuge action in the domain of low rotational frequencies, here we expand the utility of the new centrifuge design by adding to it a controlled amount of arbitrarily low angular acceleration. We outline the details of the upgraded centrifuge, dubbed ``ultraslow'' (usCFG), and the methods of its calibration. We also demonstrate the ability of usCFG to gradually spin up molecules at a rate as low as \SI{100}{MHz/ps} (i.e. three orders of magnitude slower than the conventional centrifuge) by measuring the degree of molecular alignment as the molecule follows the accelerated centrifuge rotation.
\begin{figure*}
\includegraphics[width=\textwidth]{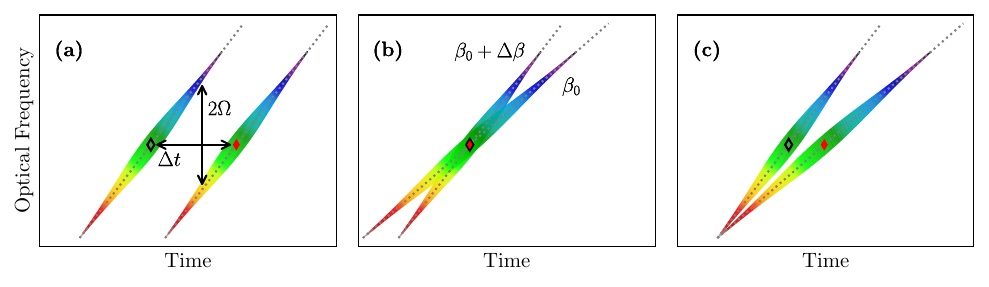}
\caption{\label{fig:slants}Time-frequency spectrograms of the two ``centrifuge arms'' for (\textbf{a}) a constant-frequency centrifuge, (\textbf{b}) an ultraslow centrifuge with zero \textit{central} frequency, and (\textbf{c}) an ultraslow centrifuge with zero \textit{initial} frequency. Both \dt{} and \db{} have been exaggerated for illustrative purposes. The diamonds mark the central frequencies of the two chirped pulses, used to define $\Delta t$. See text for the definition of $\Omega$, $\beta _{0}$, and $\Delta \beta $.}
\end{figure*}

\section{Concept of ultraslow acceleration}
As a brief review, an optical centrifuge is a linearly polarized laser pulse, whose polarization vector rotates around the direction of light propagation. The interaction between the centrifuge field and the light-induced dipole moment in the molecule of interest, provides a torque which forces the molecule to follow the rotating polarization vector. Unlike the original centrifuge design~\cite{Karczmarek1999,Villeneuve2000}, based on the technique of femtosecond pulse shaping, to create the field of cfCFG, we sent a frequency-chirped pulse into a Michelson interferometer~\cite{Wang2025}.

When polarized with opposite senses of circular polarization, the two arms of the interferometer (hereafter referred to as the two centrifuge arms) produce a linearly polarized pulse whose polarization vector rotates with a constant frequency \fcf{}. The latter is determined by the (constant) instantaneous frequency difference between the two arms, which is a function of the time delay between them, \dt{}, as shown in Fig.~\ref{fig:slants}(\textbf{a}).

Consider the optical phase of a frequency-chirped pulse, entering the interferometer:
\begin{equation}
    \varphi(t,\beta_0)=\varphi_0+\omega_0t+\beta_0 t^2,
\end{equation}
\noindent where $\varphi_0$ is the initial phase, $\omega_0$ is the optical carrier frequency, $\beta_0$ is the chirp coefficient, and higher-order dispersion has been ignored. The instantaneous optical frequency of the pulse is then
\begin{equation}
    \omega(t, \beta_0) = \varphi'(t) = \omega_0 + 2\beta_0 t.
\end{equation}
\noindent The angular frequency of the polarization vector is half the difference in optical frequencies between the two arms. After delaying one centrifuge arm with respect to the other by $\Delta t$, their interference results in the constant angular frequency of cfCFG,

\begin{align}
\label{Eq:cfCFG_Omega}
    \Omega_\mathrm{cf} &= \frac{1}{2} \left[\omega\left(t + \frac{\Delta t}{2},\beta_0\right) - \omega\left(t - \frac{\Delta t}{2},\beta_0\right)\right] \nonumber\\
    & = \beta_0 \Delta t.
\end{align}

To introduce constant angular acceleration to the centrifuge field, one must arrange for a linear increase of the instantaneous frequency difference between the two arms with time. We achieve this by adding an extra amount of frequency chirp, \db{}, to one of the two pulses in the interferometer, using a standard grating-based pulse compressor (as described in detail later in the text). As a result, the angular frequency of the polarization vector becomes

\begin{align}
    \Omega_\mathrm{us}(t) &= \frac{1}{2} \left[\omega\left(t + \frac{\Delta t}{2},\beta_0+\db\right) - \omega\left(t - \frac{\Delta t}{2},\beta_0\right)\right] \nonumber\\
    &\approx \beta_0\dt{} + \db{}t,
    \label{Eq:usCFG_Omega}
\end{align}

\noindent where we neglected the term $\db\dt/2$, since $\vert\db\vert \ll \vert\beta_0|$. Eq.~(\ref{Eq:usCFG_Omega}) above describes an ultraslow optical centrifuge with constant angular acceleration $\dot{\Omega}_\mathrm{us} =\Delta \beta$.


The time-frequency spectrograms of the two centrifuge arms with different frequency chirp rates are shown in Fig.~\ref{fig:slants}(\textbf{b}), with \dt{} set to 0. In this configuration, the rotation initially decelerates, passing through zero frequency at $t=0$, and then accelerates in the opposite direction.

By changing \dt{} and \db{}, one gains control over both the initial and the terminal frequency of the centrifuge. Fig.~\ref{fig:slants}(\textbf{{c}}) depicts the situation in which an ultraslow optical centrifuge operates similar to the one generated by means of the conventional design, i.e. accelerating from zero to a pre-defined target rotational frequency.

The time-dependent rotational frequency of the centrifuge can be expressed as:
\begin{equation}
\label{Eq:fcfg}
    \fus(t) = \frac{\Omega_\mathrm{us}(t)}{2\pi}=f_0 + \frac{\Delta\beta}{2\pi} t,
\end{equation}

\noindent where
\begin{equation}
    f_0 \equiv \fus(0) = \frac{\beta_0\Delta t}{2\pi}
    \label{Eq:central_freq}
\end{equation}
\noindent is the central frequency and the frequency window accessible by the ultraslow centrifuge, i.e. its total frequency bandwidth, is equal to
\begin{equation}
    \Delta \fus = \frac{\Delta\beta\tau}{2\pi},
    \label{Eq:freq_bandwidth}
\end{equation}
\noindent where $\tau$ is the initial pulse duration.

\begin{figure}
\includegraphics[width=\columnwidth]{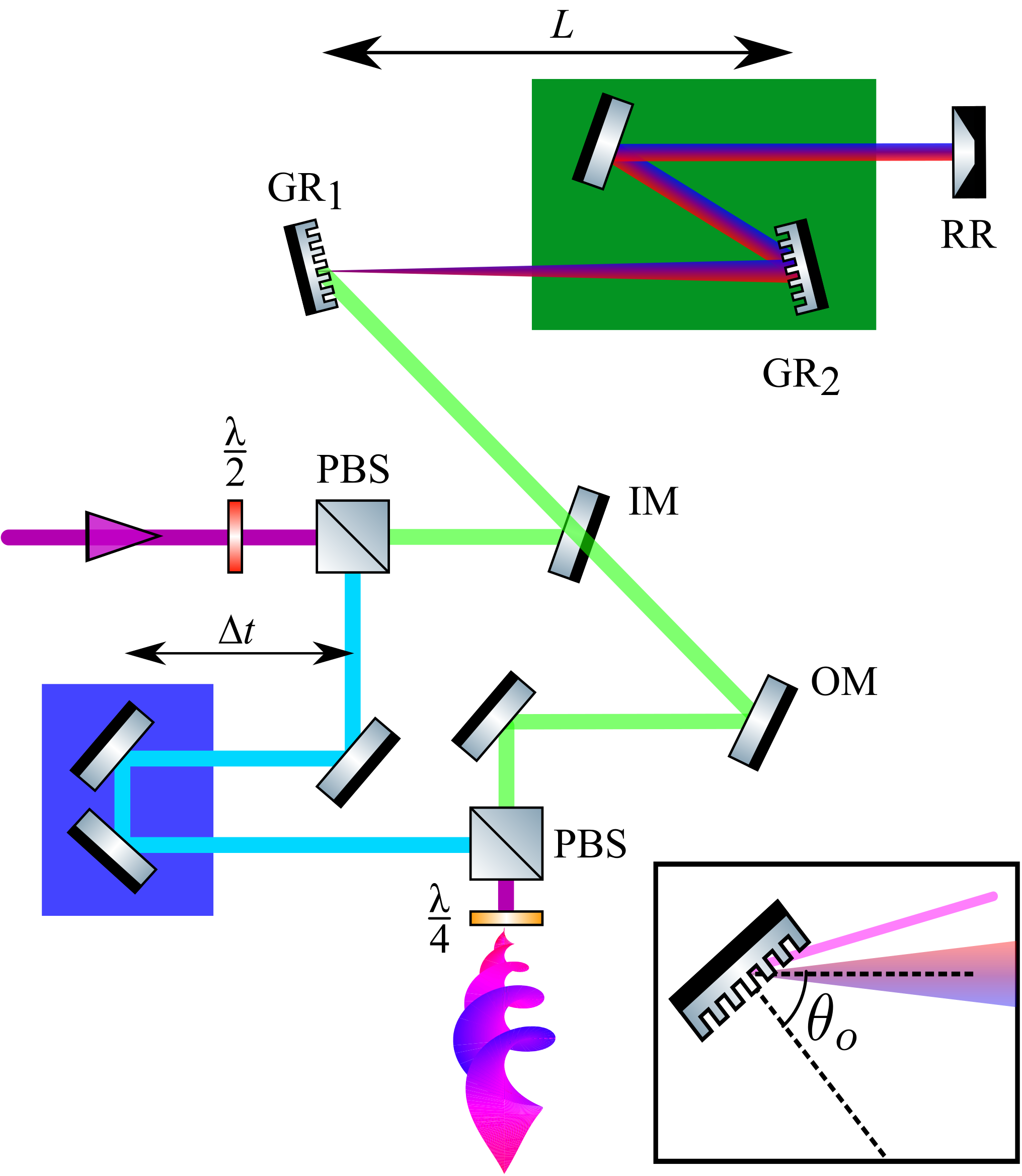}
\caption{\label{fig:shaper}
Ultraslow centrifuge pulse shaper. The chirped input pulse (purple) is split into two arms with a half-wave plate ($\lambda/2$) and a polarizing beam splitter (PBS). The transmitted ``compressor arm'' (green beam) is sent through the double-grating pulse compressor with a variable distance $L$ between the gratings ($\mathrm{GR}_1$ and $\mathrm{GR}_2$), controlled with a translation stage (green rectangle), and is vertically displaced with a retro-reflector (RR). Another translation stage (blue rectangle) controls the delay $\Delta t$ of the reflected ``delay arm'' (cyan beam). The two arms are combined on the output PBS. IM/OM mark the input/output mirrors of the compressor, respectively. A quarter-wave plate ($\lambda/4$) converts the two orthogonal linear polarizations into the circular polarizations of opposite handedness, to form the field of the centrifuge (corkscrew shape). The inset shows our definition of the grating orientation angle $\theta_0$, used in the main text.}
\end{figure}

\section{Ultraslow centrifuge shaper}
The design of the usCFG pulse shaper is shown in Fig.~\ref{fig:shaper}. The chirp difference between the two centrifuge arms is introduced by inserting a grating pair pulse compressor into one arm of the Michelson interferometer. Changing distance $L$ between the two gratings by means of a translation stage (green rectangle), results in the following approximate (for $\vert\db\vert \ll \vert\beta_0|$) change in the chirp rate~\cite{Treacy1969}:
\begin{equation}
    \db(L) \approx \frac{16\beta_0^2\pi^2c}{d^2\omega_0^3\cos^2\theta_0}L,
\label{Eq:chirp}
\end{equation}

\noindent where $1/d$ is the groove density of the gratings, $\omega_0$ is the carrier frequency of the input pulse, $\theta_0$ is the grating orientation angle as defined in the inset of Fig.~\ref{fig:shaper}, and $c$ is the speed of light. The design of the compressor arm enables us to change $L$ without altering the optical path length at the carrier frequency, thereby preserving \dt{}. Thus, the two key parameters of usCFG, its central rotation frequency $f_0$ and its frequency bandwidth \dfus{}, can be varied independently from one another.

As a relevant example of usCFG parameters, consider the centrifuge, which would accelerate from $0$ to \SI{30}{GHz}. This implies a central frequency $f_0 = \SI{15}{\giga \hertz}$ and a bandwidth $\Delta \fus = \SI{30}{\giga \hertz}$. From the parameters of our laser pulses, their central wavelength $\lambda_0= \SI{802}{\nano \metre}$, spectral width $\Delta \lambda = \SI{8.2}{nm}$ and pulse length $\tau \approx \SI{320}{\pico \second}$ (both expressed as full width at half maximum), the initial chirp coefficient can be estimated as
\begin{equation}\label{Eq:beta0}
    \beta _0 \approx \frac{\pi c \Delta \lambda }{\lambda _0^2 \tau } = \SI{3.7e-2}{rad\, ps^{-2}}.
\end{equation}
\noindent Inserting $\beta_0$ into Eqs.~(\ref{Eq:central_freq}) and (\ref{Eq:freq_bandwidth}), we arrive at the required $\dt{}=\SI{2.5}{ps}$ and $\db{}=\SI{5.9e-4}{rad\, ps^{-2}}$. Plugging this into Eq.~\ref{Eq:chirp}, together with the parameters of our gratings ($1/d=1500$~grooves/mm and $\theta_0=\SI{15}{\degree}$), results in the easy-to-implement grating separation $L= \SI{47.5}{\milli \metre}$.

\section{Calibration and tuning techniques}
To calibrate usCFG, we take advantage of the interferometric stability of the centrifuge pulse shaper (unattainable with a conventional centrifuge), which means that the orientation of the corkscrew centrifuge field, i.e. the angle of its polarization vector at any given time, does not change from one laser pulse to the other. This enables us to map out the rotation of the polarization vector in the time domain using optical cross-correlation.

\begin{figure}
\includegraphics[width=\columnwidth]{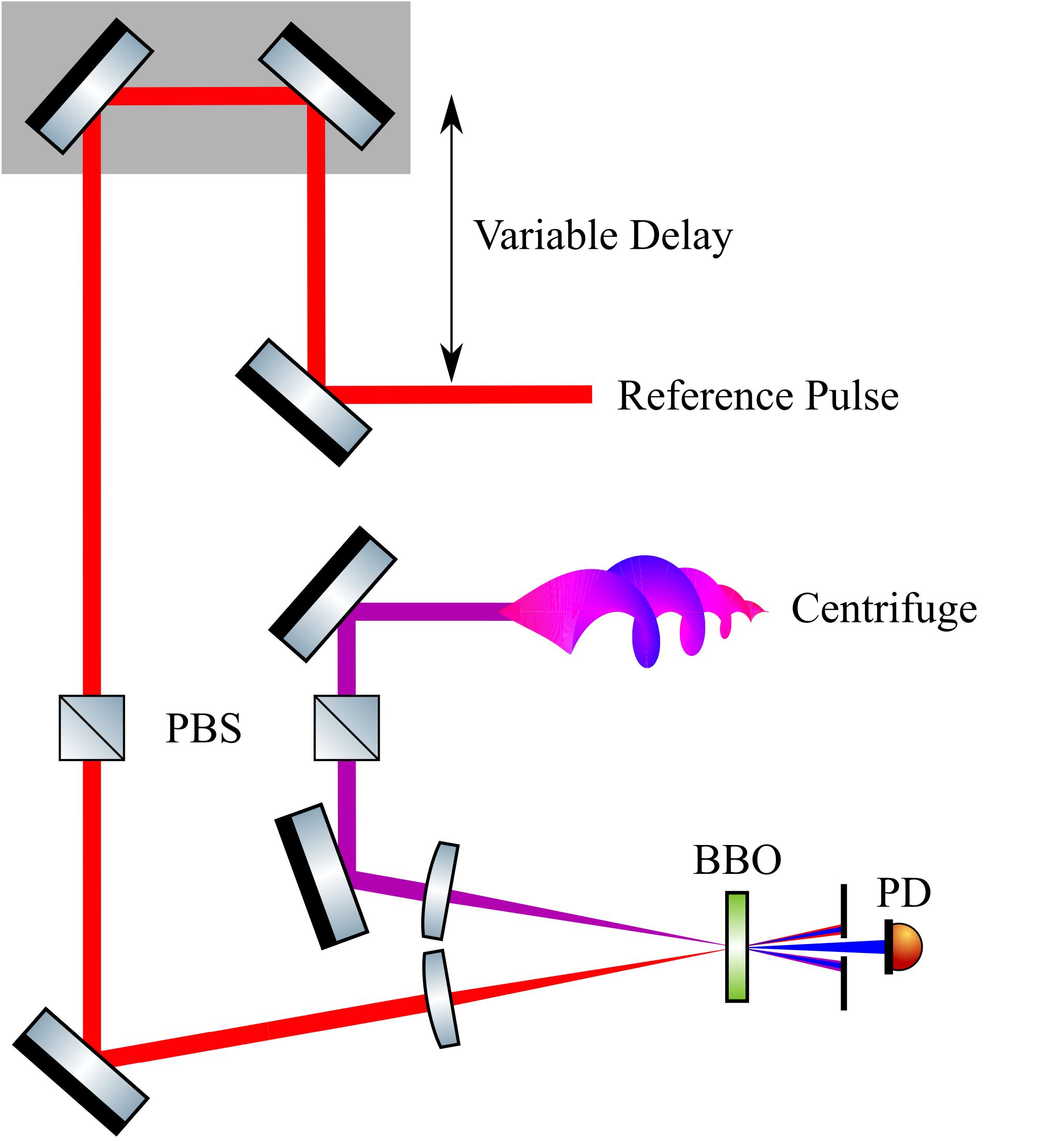}
\caption{\label{fig:xcorr_setup} Experimental layout for the cross-correlation characterization of the centrifuge field. The centrifuge pulse is passed through a linear polarizer (PBS), to project the centrifuge field vector onto the axis of a nonlinear barium borate (BBO) crystal, where it is overlapped with the linearly polarized short reference pulse. The pulses intersect at a slight angle, allowing the collection of the sum frequency generation signal with a photodiode (PD), after blocking the second harmonic generation signals from each individual pulse with an aperture. The delay time between the centrifuge and reference pulses is controlled with a variable delay stage.}
\end{figure}

The cross-correlation is performed by projecting the centrifuge field onto a single polarization axis with a linear polarizer, and then mixing it with a much shorter (\SI{\approx 120}{\femto \second}) reference pulse of the same polarization on a type-I nonlinear crystal (barium borate, BBO), as depicted in Fig.~\ref{fig:xcorr_setup}. To determine the usCFG polarization angle as a function of time, we scan the delay between the reference and the centrifuge pulse, while recording the sum-frequency generation (SFG) signal. As the latter is proportional to the square of the centrifuge field projection, the SFG signal is expected to oscillate at twice $\fus{}(t)$.

\begin{figure}
\includegraphics[width=\columnwidth]{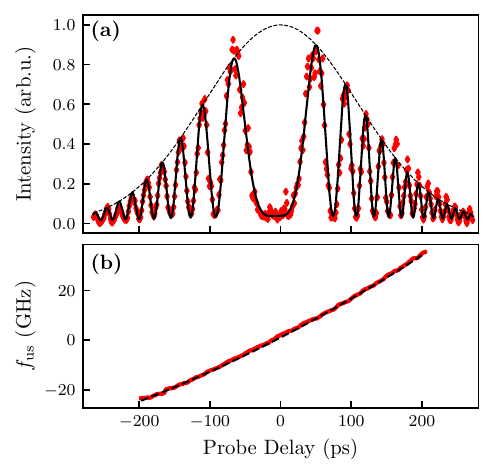}
\caption{\label{fig:xcorr_sample}
(\textbf{a}) Sample cross-correlation trace of usCFG (red diamonds), and its reconstruction from the numerically extracted instantaneous frequency (solid black). The dashed black line outlines the pulse envelope. (\textbf{b}) Extracted instantaneous frequency smoothed with a moving average (red). The dashed black curve shows $\fus(t)$, modeled to minimize the reconstruction error of the cross-correlation trace from the solid black line (\textbf{a}), with quadratic time dependence to account for third-order dispersion (see text for details). }
\end{figure}
A sample cross-correlation trace is shown in Fig.~\ref{fig:xcorr_sample}(\textbf{a}) for the centrifuge with $f_0\approx\SI{0}{GHz}$ and $\dfus \approx \SI{40}{GHz}$. The instantaneous projection of the rotating polarization vector onto the BBO axis (filtered out by the polarizing cube, PBS in Fig.~\ref{fig:xcorr_setup}) changes in time. Hence, as the usCFG-reference delay is varied, the SFG signal peaks when the centrifuge polarization is parallel to the BBO axis, and dips when they are orthogonal, which results in the observed oscillations.

The rotation decelerates in the first half of the pulse, reaching zero frequency at time zero, and then accelerates in the opposite direction. We extract the absolute value of the instantaneous frequencies of rotation by calculating the Hilbert transform of the cross-correlation signal, and differentiating the phase of the complex-valued result with respect to time. Negative sign is applied to the frequencies on the left of the inflection point, if the latter is detected. After smoothing the extracted frequency-vs-time dependence, we use it as an initial guess for fitting the signal with a frequency-chirped squared sinusoid oscillating under a Gaussian envelope. Time $t=0$ is assigned to the peak of that envelope. The corresponding rotational frequency of the centrifuge (equal to half the oscillation frequency of the SFG signal) is plotted in Fig.~\ref{fig:xcorr_sample}(\textbf{b}), showing near-constant acceleration across the centrifuge profile.

A slight increase in the rotational acceleration with time is apparent in Fig.~\ref{fig:xcorr_sample}(\textbf{b}) and can be attributed to positive third-order dispersion (TOD) in the initial chirped pulses from the laser system ($\gamma _0$; the effect of this TOD on the centrifuge acceleration has been analyzed in detail in Ref.~\citenum{Wang2025}) and that added by the grating pair, $\Delta \gamma $. Owing to such TOD, the optical phase is varying in time as
\begin{equation}
\label{Eq:TOD}
\varphi(t, \beta, \gamma) = \varphi_0 + \omega_0 t + \beta t^2 + \gamma t^3,
\end{equation}
\noindent where $\beta = \beta _0,\, \gamma = \gamma _0$ for the delay arm of the centrifuge, and $\beta = \beta _0+\db,\, \gamma = \gamma _0+\dg$ for the compressor arm. This temporal phase results in the following instantaneous centrifuge angular frequency [cf. Eq.~(\ref{Eq:usCFG_Omega})]:
\begin{equation}
    \Omega_\mathrm{us}(t) \approx \beta_0\Delta t+ \big(\db+3\gamma _0\dt\big) t+ \frac{3}{2} \dg t^2,
    \label{Eq:usCFG_Omega_TOD}
\end{equation}
\noindent where we assumed $\dg\ll \gamma _0$. The last term in the above expression explains the small quadratic nonlinearity in the time dependence of the usCFG frequency, visible in Fig.~\ref{fig:xcorr_sample}(\textbf{b}).

\begin{figure*}
\includegraphics[width=\textwidth]{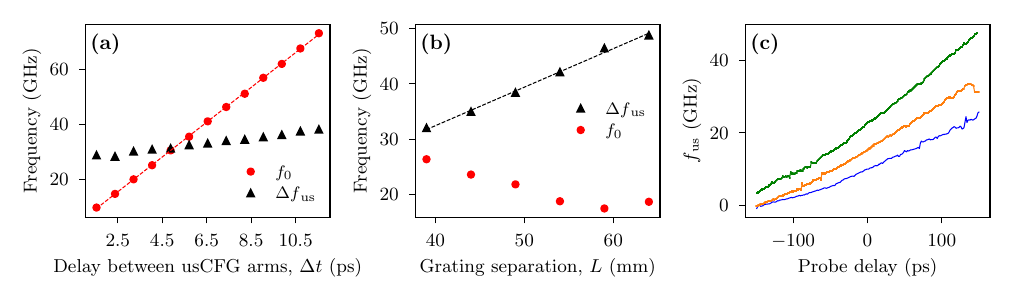}
\caption{\label{fig:xcorr_calibration} Tunable behaviour of usCFG, characterized by means of optical cross-correlation. (\textbf{a}) Central frequency of usCFG, $f_0$, at different delays between the centrifuge arms, $\Delta t$. The grating separation in the compressor was recorded with a fixed $L \approx \SI{39}{\milli\metre}$. (\textbf{b}) The frequency bandwidth of usCFG, \dfus{}, at different grating separations $L$, recorded with a fixed value of $\Delta t \approx \SI{3}{\pico\second}$. (\textbf{c}) Extracted $\fus(t)$ smoothed with a moving average for three different centrifuges, tuned to have terminal frequencies of approximately \SI{20}{\giga \hertz} (blue), \SI{30}{\giga \hertz} (orange), and \SI{40}{\giga \hertz} (green).}
\end{figure*}

To illustrate the tunability of the \textit{central frequency} of the ultraslow centrifuge, we performed cross-correlation measurements for various values of \dt{}, while keeping $L$ constant. Fig.~\ref{fig:xcorr_calibration}(\textbf{a}) shows the expected linear relationship between $f_0$ and $\Delta t$, as predicted by Eq.~(\ref{Eq:central_freq}), and the desired minimal changes in $\Delta \fus{}$. The slope $f_0/\dt = \beta _0/2\pi = \SI{6.30(2)}{\giga \hertz \per \pico \second}$ is similar to the value of \SI{6.4(2)}{\giga \hertz \per \pico \second} reported in Ref~\cite{Wang2025} for a cfCFG constructed from the same laser source with the same $\beta _0$.

To demonstrate our ability to vary the \textit{acceleration} of usCFG, Fig.~\ref{fig:xcorr_calibration}(b) shows the extracted values of $\Delta \fus{}$ at different values of the grating separation $L$, recorded with fixed delay stage position. Here again, we observe a linear dependence, as anticipated from Equations~(\ref{Eq:freq_bandwidth}) and~(\ref{Eq:chirp}). The slope of \SI{0.70(2)}{\giga \hertz \per \milli \metre} is in good agreement with the calculated value of \SI{0.71(1)}{\giga \hertz \per \milli \metre}. Owing to the imperfections of the beam alignment in the centrifuge shaper, slight unintentional changes in the delay between the two interferometer arms always accompany the deliberate tuning of the grating separation, resulting in small variations of $f_0$.

Despite the nonlinear effects of TOD on $\fus(t)$ and the calibration characteristics of usCFG, tuning the centrifuge is readily accessible by appropriately adjusting both $\Delta t$ and $L$ (and thus $f_0$ and $\Delta \fus{}$). In Fig.~\ref{fig:xcorr_calibration}(\textbf{c}), we demonstrate the performances of three different centrifuges that all start from around zero frequency at the rising edge of the pulse, but reach different terminal frequencies by its falling edge.

\section{Proof of operation in the molecular jet}
\begin{figure}
\includegraphics[width=0.8\columnwidth]{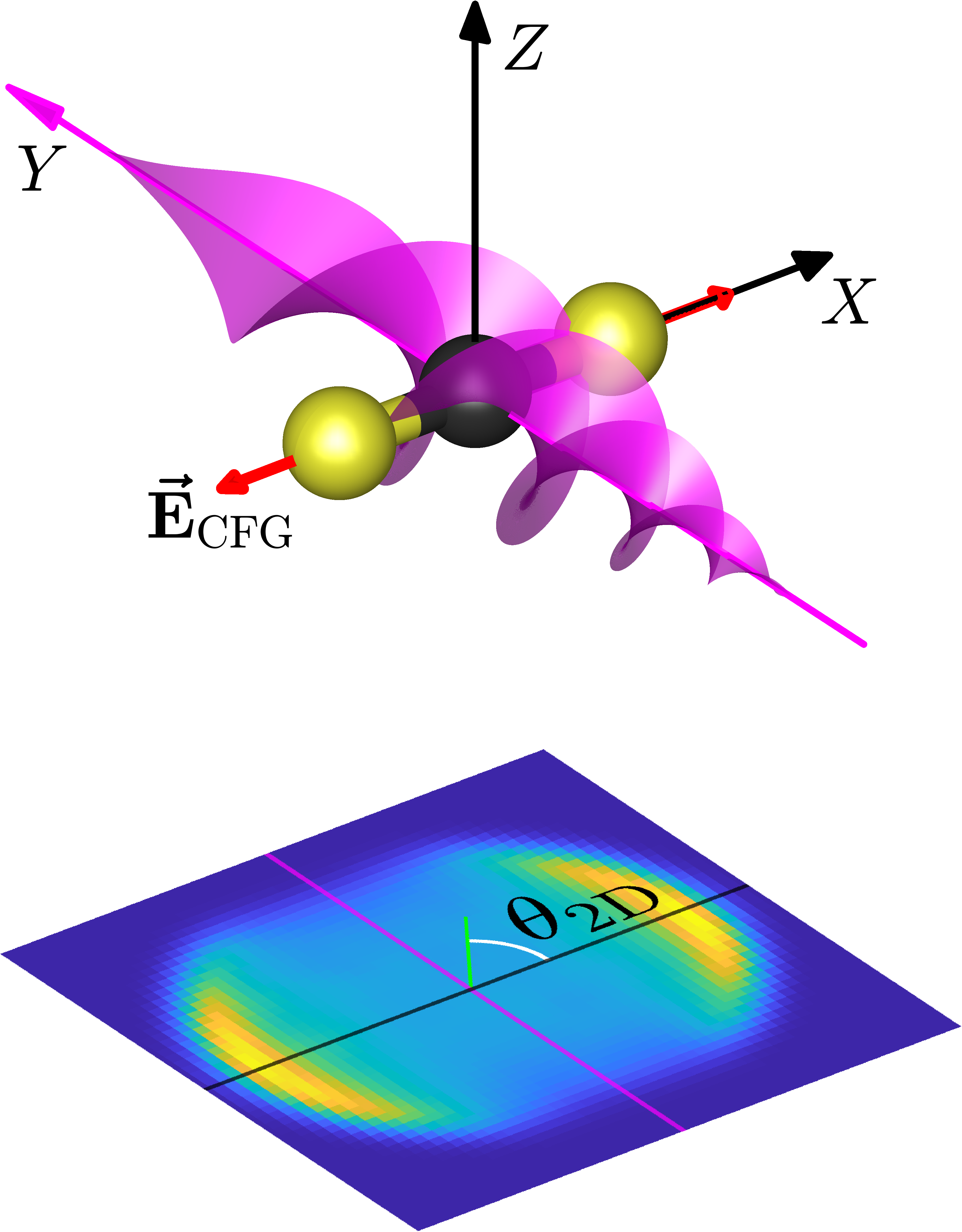}
\caption{ \label{fig:VMI_geometry}
Geometry of the Velocity Map Imaging experiment. The magenta corkscrews represent the electric field of the centrifuge ($\vec{\mathbf{E}}_\mathrm{CFG}$), propagating along the $Y$ axis. The VMI setup is oriented along the $Z$ axis, projecting a two-dimensional velocity map image of ion fragments onto the detector, shown below the $XY$ plane. $\theta_{\text{2D}}$ is the angle of a detected ion fragment, measured with respect to the $XZ$ plane. A \CSS{} molecule is shown aligned to $\vec{\mathbf{E}}_\mathrm{CFG}$. }
\end{figure}

To demonstrate the utility of usCFG to induce molecular rotation, we employed the well-known technique of Velocity Map Imaging (VMI)~\cite{Eppink1997}. Centrifuge pulses (\SI{\approx 330}{\pico \second}, \SI{3}{\milli \joule} pulse energy) are focused onto a molecular jet of carbon disulfide (\CSS) molecules, which are then Coulomb-exploded by probe pulses (\SI{\approx 120}{\femto \second}, \SI{150}{\micro \joule} pulse energy) linearly polarized in the direction perpendicular to the plane of the VMI detector (axis Z in Fig.~\ref{fig:VMI_geometry}). The centrifuge pulses and probe pulses used in this experiment have a repetition rate of \SI{1}{\kilo \hertz}, a central wavelength of \SI{802}{\nano \meter}, and focused peak intensities of \SI{\approx 1e12}{\watt\per\square\cm} and \SI{\approx 4e14}{\watt\per\square\cm}, respectively. \CSS{} gas is mixed with \SI{8}{\bar} of helium at a concentration of \SI{1000}{ppm}, with an estimated rotational temperature below \SI{10}{\kelvin} after supersonic expansion through a pulsed nozzle into vacuum.

The orientation of a molecule at the time of the nearly-instantaneous Coulomb explosion is inferred from the velocities of the resultant ion fragments. By scanning the arrival time of the probe with respect to the centrifuge pulse, and measuring the 2D projection of the S$^+$ ion fragment velocity distribution, as depicted in Fig.~\ref{fig:VMI_geometry}, we observe the rotational dynamics of molecules inside the centrifuge. $\theta_\mathrm{2D}$ describes the angle between the position vector of an ion hit on the detector plane, and the rotational plane of the centrifuge ($XZ$). The alignment of the centrifuged molecules is then quantified by the average, \costh{}, over many ion hits. For each complete rotation of the centrifuge, the centrifuged molecule is expected to align with the laboratory $X$ axis (see Fig.~\ref{fig:VMI_geometry}) twice, resulting in the oscillations of \costh{}(t) at twice the centrifuge frequency \fus{}(t).

\begin{figure}
\includegraphics[width=\columnwidth]{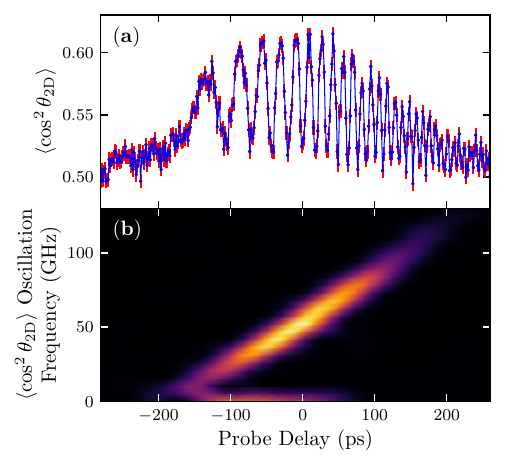}
\caption{\label{fig:Jet_scan}
(\textbf{a}) $\costh(t)$ of S$^+$ fragments (blue markers with red error bars), recorded during the forced rotation of \CSS{} molecules by usCFG. Blue lines connect the data points to guide the eye. (\textbf{b}) Spectrogram produced by applying a short-time Fourier transform (STFT) to the signal plotted in (\textbf{a}). The tilted trace accelerating up to $\sim\SI{100}{\giga \hertz}$ confirms that the molecules are following the accelerated rotation of the centrifuge. The equidistant lines are numerical artifacts related to the window size of the STFT procedure.}
\end{figure}

Fig.~\ref{fig:Jet_scan}(\textbf{a}) shows the degree of alignment of \CSS{} molecules \textit{during} the usCFG. The centrifuge was tuned to start at zero rotational frequency, i.e. $\fus{}(t\approx\SI{-150}{ps})=\SI{0}{GHz}$, and accelerate to \SI{\approx25}{GHz} at the peak of its intensity profile around $t = 0$~ps. Given the initial chirp rate $\beta_0 \approx$ \SI{40}{\radian\,\pico \second^{-2}}, this required the following parameters of the centrifuge shaper: $\Delta\beta \approx$ \SI{1e-3}{\radian\, \pico \second^{-2}}, and $\Delta t \approx$ \SI{4}{\pico \second}.

Clear oscillations of $\costh{}(t)$ confirm that \CSS{} molecules have been captured by the centrifuge and follow its rotation. As molecules spin in the $XZ$ plane, the ejection angle of S$^+$ ions periodically alternates between along the $X$-axis, resulting in $\costh > 0.5$, and along the $Z$-axis, with $\costh \approx 0.5$. Evidently, the oscillation period is decreasing with time, attesting to the rotational acceleration of molecules by usCFG. The spectrogram in Fig.~\ref{fig:Jet_scan}(\textbf{b}) was produced by applying a short-time Fourier transform to the data presented in Fig.~\ref{fig:Jet_scan}(\textbf{a}), allowing for a time-frequency map of molecular rotation.

\section{Summary}
In summary, we have developed an alternative implementation of the optical centrifuge, which enabled us to lower its rotational acceleration by three orders of magnitude, as compared to the conventional centrifuge design. We have demonstrated the ability to tune the acceleration in the range of around $\SI{100}{MHz/ps}$, and applied this ``ultraslow optical centrifuge'' to \CSS{} molecules in a molecular jet, confirming their successful capture and spinning by the centrifuge field. Using usCFG to investigate the effect of the helium environment on the rotational dynamics of molecules embedded in helium nanodroplets is the subject of current study. In that regard, the ultraslow centrifuge opens exciting new avenues to probe the interaction of rotationally excited molecules with quantum many-body systems.

\begin{acknowledgments}
This research has been supported by grants from CFI, BCKDF, and NSERC, and carried out under the auspices of the UBC Centre for Research on Ultra-Cold Systems (CRUCS).
\end{acknowledgments}

\bibliographystyle{apsrev4-2}

%

\end{document}